\documentclass[a4paper,fleqn,usenatbib]{mnras}
\usepackage{float}

\usepackage{newtxtext,newtxmath}
\usepackage[applemac]{inputenc}
\usepackage[T1]{fontenc}
\usepackage{ae,aecompl}

\usepackage{graphicx}	% Including figure files
\usepackage{amsmath}	% Advanced maths commands
\usepackage{amssymb}	% Extra maths symbols

\title[Spectral multiplexing using stacked VPHGs - Part I]{Spectral multiplexing using stacked VPHGs - Part I}

\author[A. Zanutta \& M. Landoni et al.]{
A. Zanutta$^{1}$\thanks{Contact e-mail: \href{mailto:alessio.zanutta@brera.inaf.it}{alessio.zanutta@brera.inaf.it}} \& M. Landoni$^{1}$\thanks{Contact e-mail: \href{mailto:marco.landoni@brera.inaf.it}{marco.landoni@brera.inaf.it}}, M. Riva$^{1}$,
and A. Bianco$^{1}$
\\
$^{1}$INAF - Osservatorio Astronomico di Brera, via E. Bianchi 46, 23807 Merate (LC), Italy
}

\date{Accepted XXX. Received YYY; in original form ZZZ}

\pubyear{2017}

\begin{document}
\label{firstpage}
\pagerange{\pageref{firstpage}--\pageref{lastpage}}
\maketitle

\begin{abstract}
Many focal-reducer spectrographs, currently available at state-of-the art telescopes facilities, would benefit from a simple refurbishing that could increase both the resolution and spectral range in order to cope with the progressively challenging scientific requirements but, in order to make this update appealing, it should minimize the changes in the existing structure of the instrument. In the past, many authors proposed solutions based on stacking subsequently layers of dispersive elements and record multiple spectra in one shot (multiplexing). Although this idea is promising, it brings several drawbacks and complexities that prevent the straightforward integration of a such device in a spectrograph.
Fortunately nowadays, the situation has changed dramatically thanks to the successful experience achieved through photopolymeric holographic films, used to fabricate common Volume Phase Holographic Gratings (VPHGs). Thanks to the various advantages made available by these materials in this context, we propose an innovative solution to design a stacked multiplexed VPHGs that is able to secure efficiently different spectra in a single shot. This allows to increase resolution and spectral range enabling astronomers to greatly economize their awarded time at the telescope. In this paper, we demonstrate the applicability of our solution, both in terms of expected performance and feasibility, supposing the upgrade of the Gran Telescopio CANARIAS (GTC) Optical System for Imaging and low-Intermediate-Resolution Integrated Spectroscopy (OSIRIS).
\end{abstract}


%%%%%%%%%%%%%%%%%%%%%%%%%%%%%%%%%%%%%%%%%%%%%

%%%%%%%%%%%%%%%%% BODY OF PAPER %%%%%%%%%%%%%%%%%%

\section{Introduction}
The current state-of-the-art spectroscopic facilities could be divided in two main groups depending on the resolution. The first one is characterized by a low resolution (R < 2000), particularly suitable for multi-object spectroscopy or Integral Field Unit (IFU). Among them we can find examples like ESO VIMOS \citep{fevre03}, FORS1-2 \citep{appenzeller98}, MOSFIRE \citep{mclean12}, the FOSC at ESO-NTT \citep{snodgrass08, buzzoni84} or OSIRIS at GTC \citep{cepa10}.
The second one is featured by a high resolution (R $>>$ 4000), which is guaranteed through diffractive elements like echelle or echellette grating. Among this group, successfully examples are HIRES at Keck \citep{voigt02}, ESO UVES \citep{dekker00}, CRIRES \citep{kaufl08} or the most recent ESO X-SHOOTER \citep{vernet11}.

The resolution plays a key role in the era of 10 m class telescopes since the only way to increase the sensitivity of a spectrum, in terms of detectable features and accuracy, is to increase it as much as possible while maintaining a good signal-to-noise ratio (SNR) over a wide spectral range. Current focal reducer spectrographs, like the GTC-OSIRIS, already provide diffraction gratings that allow to secure spectra with R $\geq$ 2000 but, unfortunately, their spectral range is very narrow.
Therefore, to obtain a spectrum from 4000 to 10000 $\textrm{\AA}$, it would require to observe the same source multiple times, thus wasting an enormous amount of awarded time. 
On the other hand, deciding to operate in alternative facilities based on echelle gratings (e.g. ESO-XSHOOTER or Keck-ESI), a scientist could secure spectra with wide wavelength coverage at high resolution (R $>$ 4000). Nevertheless, spectrographs like GTC-OSIRIS or EFOSC are much simpler, widely diffused in a large number of optical telescopes and with a flexible design that includes imaging capabilities. Therefore, an improvement of these facilities that allow to fill the gap between the two layouts is really attractive.

In most astrophysical topics, the increase of the resolving power of secured spectra permits to achieve many scientific advantages. For example, the use of ESO-XSHOOTER spectrograph produced a vast number of spectra of QSOs at R > 5000, allowing to better understand the physical state and chemical composition of the IGM \citep{lopez16, odorico16a, odorico16b}. 
Another interesting topic tackled with the same instrument is the determination of the redshift of BL Lac objects. These sources are active nuclei of massive elliptical galaxies whose emission is dominated by a strong non-thermal continuum \citep{falomorev, shaw13, massarorev, sandrinelli13} that prevents the determination of their distance that is, however, mandatory for constraining models of their emission or to better understand absorption of hard $\gamma$-rays by Extragalactic Background Light (EBL). In fact for example, \cite{pita14, landonixsh} demonstrated that this problem could be mitigated by securing spectra with high SNR and increased resolution to detect fainter spectral features, necessary to estimate the redshift (and thus the distance) of the source.

As already pointed out, it is possible to collect spectra of astrophysical sources at high resolution (R > 4000) with focal reducer instruments but the price to pay is a limited spectral range. A clever solution could be the substitution of the single high R element, whose range is limited, with a new dispersive device capable to increase spectral coverage, simultaneously recording multiple high R spectra in different ranges (multiplexing). Being able to deliver such dispersive element will achieve a great improvement of the expected throughput of already commissioned spectrographs. In this paper, we focus on the GTC-OSIRIS as a candidate and demonstrator for housing our innovative device. In particular, we report on two different cases. The first one allows to secure two spectra in a single shot increasing the resolution by a factor of $\sim$ 2 while the second challenges instruments like X-SHOOTER combining three high resolution spectral regions simultaneously.
This work is organized as follows: in Section \ref{sec:th} we report on the theoretical background and the design principle of these multiplexed devices. In Section \ref{sec:simu} we discuss the expected performances on the sky through simulations, while in Section \ref{sec:conclusions} we give our conclusions.

\section{Grating Multiplexing -- theory and applicability}\label{sec:th}
\begin{figure}
\begin{center}
	\includegraphics[width=0.8 \columnwidth]{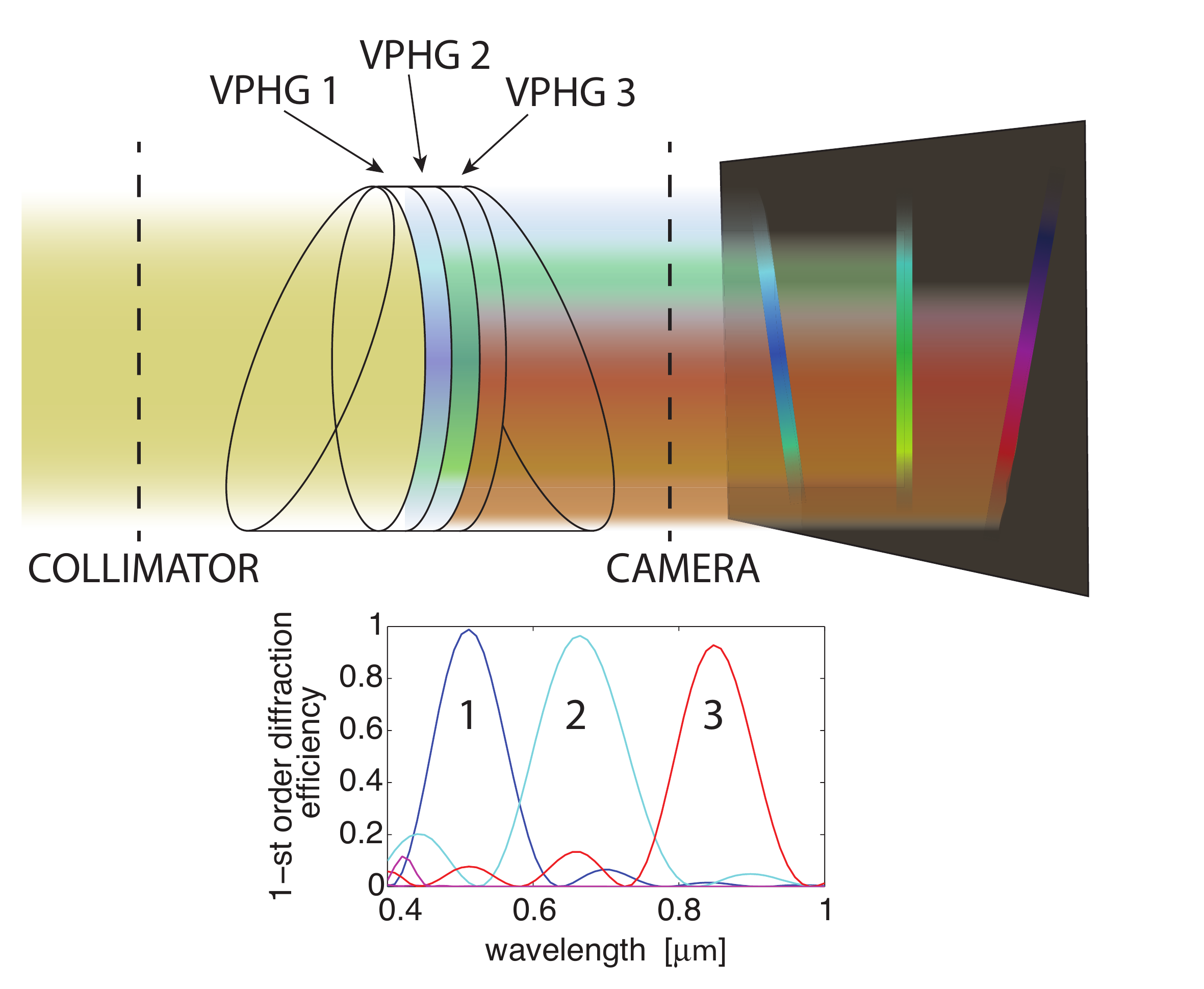}
    \caption[scheme1]{Scheme of a possible application of a multiplexed device in GRISM\footnotemark{} mode. Multiple and high dispersive VPHG layers compose the multiplexed element which produces on the CCD, spectra of the slit in different spatial locations (one for each grating layer). In the inset are reported the possible uncombined efficiencies, peaked in different spectral ranges.}
    \label{fig:multiplex_scheme2}
\end{center}
\end{figure}

As highlighted in the introduction, being able to simultaneously record multiple spectra of wavelength ranges, or alternatively, to have a high resolution element that covers a very wide spectral range, brings a huge advantage to the astronomical community. 
In particular, depending on the optical layout, a spectrograph would benefit of the possibility to 
increase the resolution or the spectral range, maintaining the same exposure times.
Otherwise, a typical focal reducer imager, like FORS or OSIRIS, would benefit from the combination of very low dispersion gratings to acquire multiple snapshots of the same field in different bands simultaneously, as depicted in the cartoon of Figure \ref{fig:multiplex_scheme2}.

In the present study we have tried to answer to these needs, testing the feasibility of a new type of dispersive element, that can result in a huge technological boost for those instruments that are becoming obsolete and for the new ones that are yet to be built.

\begin{figure}\begin{center}
	\includegraphics[width=0.8 \columnwidth]{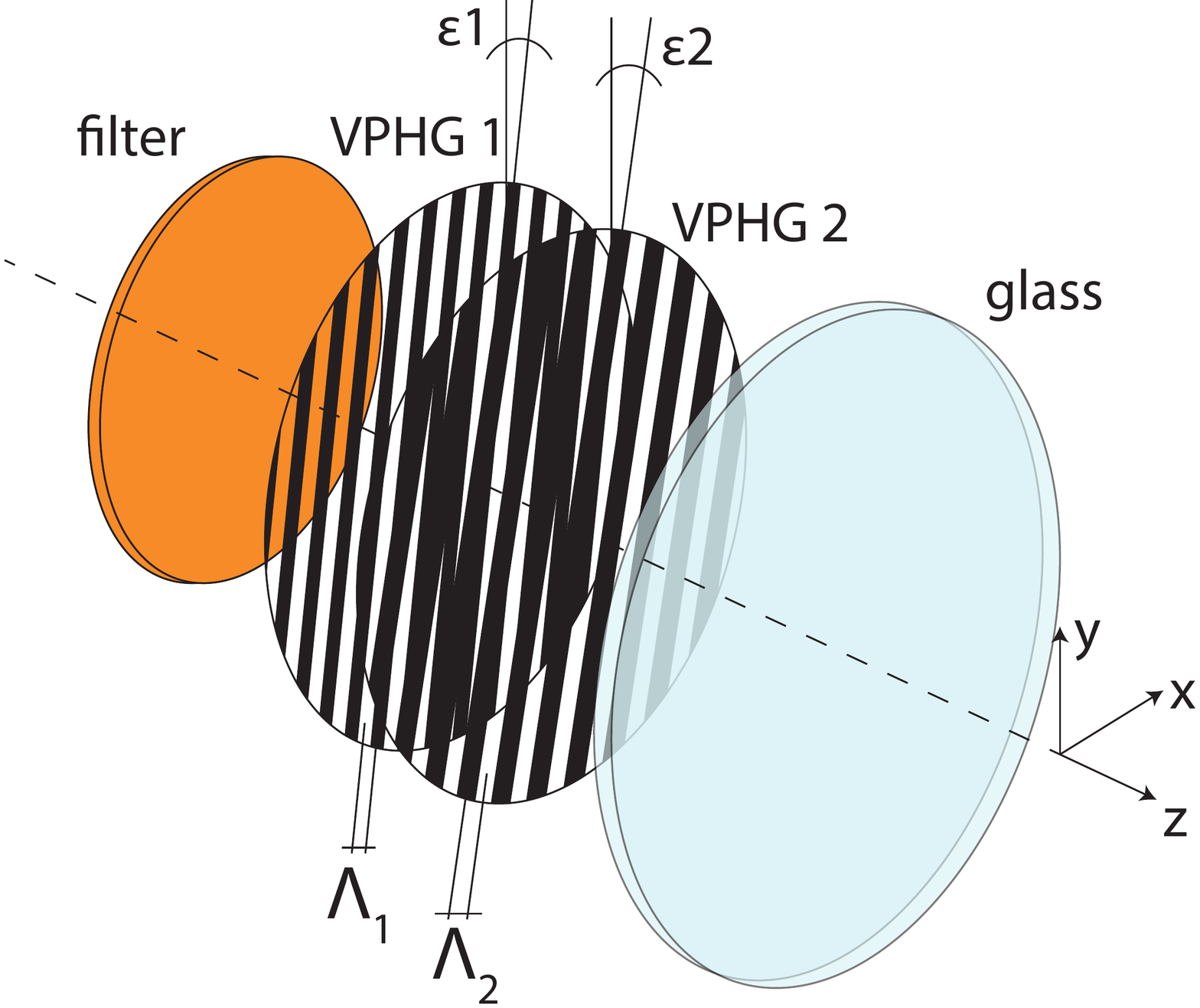}
    \caption{Scheme of the layers composing a multiplexed grating. VPHG1 and VPHG2 may have different line density and orientation (clock) in order to separate the two spectra along the y direction. Filter and glass are not mandatory elements and can be replaced with other layers such as prisms.}
    \label{fig:multiplex_scheme1}
\end{center}\end{figure}

The general idea is to place multiple gratings (multiplexed), stacked subsequently, in a way that they will produce simultaneously spectra of different wavelength regions.
The basic concept of a transmissive element is sketched in Figure \ref{fig:multiplex_scheme1}.
%%%our transmission multiplexed element would mainly consist of multiple gratings, stacked in one single optical element.

Each spectrum in the instrument's detector is designed to cover a specific wavelength range, according to the scientific case that has to be studied. Consequently, the design phase is indeed a crucial part in the definition of the characteristics of the multiplexed dispersive element. Moreover, strategies to separate the  spectra avoiding their overlapping should be considered.

In this particular configuration, since the grating layers are superimposed, the key idea is to apply a small rotation along optical axis ($\varepsilon$ in Figure \ref{fig:multiplex_scheme1}) between the layers, in order to separate (along the y direction) the different spectra appearing on the detector.

%Being able to secure multiple spectra with one exposition, allows to extend the analyzed spectral range (maintaining the same resolution R) or to increase significantly the resolution of the system in the same spectral range. This system is therefore suitable for upgrading an already built instrument, giving a great enhancement by simply replacing the dispersive element.
Being able to secure multiple spectra with one exposure, the analyzed spectral range is extended (maintaining the same resolution R) or the resolution of the system in the same spectral range is significantly increased. This system is therefore suitable for upgrading an already built instrument, giving a great enhancement by a simple replacement of the dispersive element, which preserves all the existing abilities (\textit{e.g.} the imaging in FOSC).

\footnotetext{A GRISM is a combination of a prism and grating arranged so that light at a chosen central wavelength passes straight through. The advantage of this arrangement is that the same camera can be used both for imaging (without the grism) and spectroscopy (with the grism) without having to be moved. Grisms are inserted into a camera beam that is already collimated. They then create a dispersed spectrum centered on the object's location in the camera's field of view.}

The type of dispersive element that we have considered in this study is the transmissive Volume Phase Holographic Grating, VPHG \citep{Barden1998}. They consist in a periodic modulation of the refractive index ($\Delta$n) in a thin layer of a photosensitive material. These elements represent today the most used dispersive elements in astronomy and yet the element whose performances are most difficultly surpassed in both low and medium resolution spectrographs \citep{spano2006challenges, baldry2004volume, pazder2008vph}.

Since many different VPHGs are usually integrated inside astronomical spectrographs and each of them is a custom designed grating, each astronomical observation can take advantage of specific dispersive elements with features tailored for achieving the best performances. Accordingly, the design and manufacturing of highly efficient and reliable VPHGs require photosensitive materials where it is possible to control both the refractive index modulation and the film thickness $d$, in order to tune the device's efficiency. 

Regarding the holographic materials, up to now Dichromated Gelatins (DCG) is considered the reference material thanks to the very large modulation of the refractive index that can be stored \citep{Xiong1998, Bianco2012}, which turns into relative large bandwidth in high dispersion gratings. Unfortunately this material requires a complex chemical developing process making it difficult for large scale and large size production.
Moreover, the material is sensitive to humidity, therefore, it is necessary to cover the grating with a second substrate, burdening the control of the wavefront error.
 
The availability of holographic materials with similar performances, but with self-developing properties is desirable, because they will not require any chemical post-process and moreover, the $\Delta$n formation could be monitored and set during the writing step. 

Photopolymers are a promising class of holographic materials and today, they are probably the best alternative to DCG, thanks to the improved features in terms of refractive index modulation, thickness control and dimension stability \citep{Lawrence2001, Bruder2011, Ortuno2013, Fernandez2015}. A lot of studies have been carried out to understand deeply the behavior of this class of materials. Moreover, the formation of the refractive index modulation has been recently studied \citep{Gleeson2009, Gleeson2011, Li2014a, Li2014b}, through the development of models that predict the trends as function of the material properties and writing conditions \citep{kowalski2016design}.

We already demonstrated in other papers the use of photopolymers for making astronomical VPHGs with performances comparable to those provided by VPHGs based on DCGs and good aging performances \citep{zanutta14aging}, but with a much simpler production process \citep{Zanutta2016}. Therefore, we think that the big advantages of this novel holographic material could be the key point to realize the multiplexed dispersive element.

The newly \citep{Bruder2010, Berneth2011, Berneth2013} developed photopolymer film technology (Bayfol HX$^{\circledR}$ film) evolved from efforts in holographic data storage (HDS) \citep{Dhar2008} where any forms of post processing is unacceptable. These new instant developing recording media open up new opportunities to create diffractive optics and have proven to be able to record predictable and reproducible optical properties \citep{Bruder2009}.
Depending on the application requirements, the photopolymer layer can be designed towards e.g. (high or low) index modulation, transparency, wavelength sensitivity (monochromatic or RGB) and thickness to match the grating's wavelength and/or angular selectivity.

Since the material consists in the holographic layer coupled with a polymeric substrate with a total thickness of ca. 60 - 150 $\mu$m,
%Since this material can be realized in thin films (common thickness of one layer is ca. 60 - 150 $\mu$m) and the glass window serves only as a rigid substrate, they possess a further advantage: 
it can be laminated or deposited one on top of the other after having been recorded,
making straightforward the stacking realization.
%enhancing the easiness of the manufacturing process. 

Clearly, another possibility is to holographically record multiple gratings inside the same layer but, as described later, in order to optimize the efficiency curves, usually very different thicknesses  are required for each grating, therefore this strategy will not let us have the advantage to tune the response curves in the design process.

%%%%%
\subsection{Working principle}
As stated at the beginning of this section, the design concept consists in placing a set of transmission VPHGs stacked subsequently (multiplexed) (see Figure \ref{fig:multiplex_scheme1}).
As highlighted in the figures, this device will form one single optical element whose dimensions are comparable to standard VPHGs already available in the target instrumentation.

Some attempts have been made to explore this idea \citep{muslimov2016moderate, battey1996spectrograph} but, although steps have been made in the right direction, the proposed solutions are limited by the necessity of a newly designed spectrograph, and do not take into consideration the crucial efficiency optimization that, without proper design, will make the device ineffective. 

Hence, to preserve integration simplicity, one has to mix materials, design strategies and required performances, in order to produce multiplexed dispersive elements that could be easily integrated in an available instrument.
This gives to astronomers the possibility to enhance the resolution (and spectral coverage) by simply replacing the disperser already installed in the optical path.

%Spectrographs like OSIRIS, EFOSC or the Goodman at CTIO, characterized by medium and low-dispersion, would directly benefit by the installation of these devices without changing any optical component beside the diffraction grating. Such instruments are widely spread across telescopes of any aperture, therefore there is the strong need to find a reliable technological solution that can ensure a low-cost refurbishment and brings new scientific life to these facilities, that would be otherwise decommissioned. In this scenario the realization of a new spectrograph based on multiplexed grating is therefore simpler than upgrading an existing one. In this case, the realization of a multiplexed device, with common materials (for astronomical VPHGs) and with the spectroscopic capabilities highlighted before, would be very difficult since the compactness could not be achieved and the manufacturing reliability are prohibitive.

Regarding the material, thanks to the crucial capability to finely tune the refractive index modulation $\Delta$n \citep{zanutta2016photopolymer} and the slenderness of the film containing the grating, Bayfol HX$^{\circledR}$ photopolymers by Covestro gave us the possibility to design the multiplexing element to: 

\begin{enumerate}
\item realize a compact and thin device that can be integrated as replacement in many already existing instruments \citep{aj2016, Zanutta2014};
\item tune the single stacked efficiency in such a precise way that they will not interfere with each other and obviate to all the problems related to the realization of these devices;
\item match the design requirements and obtain high efficiency;
\item stack multiple layers of gratings in one single device for the simultaneous acquisition of multiple spectra with a broad wavelength band.
\end{enumerate}

In the multiplexed device, each layer will generate a portion of the spectrum that all together will compose the total dispersed range required.
Such pieces, on the detector, will be disposed one on the top of each other, resulting in a total spectral range that is far wider than the one obtainable using a single grating with a comparable dispersion.

\subsection{Issues in details: the geometrical effect}
Although the stack of subsequent diffraction elements brings many advantages, some constraints and critical points arise and should be discussed. 

The first one is purely a geometrical effect and is related to the propagation of the incoming beam throughout multiple dispersing elements that must not interfere with each other.
For this reason, it has to be taken into account that the incoming beam is diffracted multiple times (since it encounters two or more dispersive elements) according to the grating equation

\begin{equation}
m \lambda \Lambda = \sin \alpha + \sin \beta
\label{eq:grating}
\end{equation}
with $m$ the number of the diffracted order, $\Lambda$ the line density of the VPHG and $\alpha$, $\beta$ the incidence and diffracted angles respectively. \\

\begin{figure}
\begin{center}
	\includegraphics[width=1.05 \columnwidth]{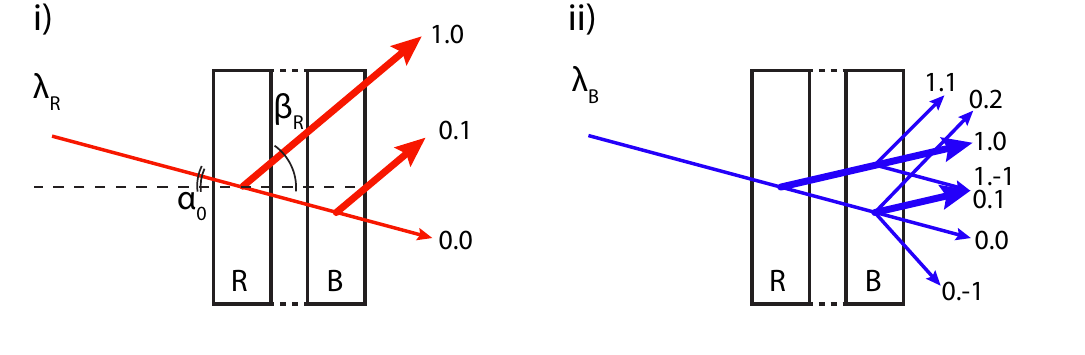}
    \caption{Monochromatic beam propagation in a two-multiplexed device (i - red wavelength case, ii - blue wavelength case). In the beam notation J.K, the number J is the diffraction order of the first grating and K the one of the second grating. R identifies the grating layer designed for dispersing efficiently the red wavelengths, while B the blue ones. Bold lines are the orders that we want to exploit in the detector}
    \label{fig:stack_scheme}
\end{center}
\end{figure}

Fixing $\lambda$, different combinations of diffraction orders can occur, resulting in light diffracted in different directions.
%Fixing $\lambda$, $\alpha$ and $\Lambda$, different diffraction orders appear out from the grating, each one with different directions.
In particular let us consider for simplicity an example of two stacked multiplexed elements as shown in Figure \ref{fig:stack_scheme}. Each grating is optimized for dispersing efficiently a specific wavelength range (labelled B for blue and R for red). Let's suppose that the two possess the same line density.
If a red monochromatic wavelength $\lambda_{R}$ (case i) enters the multiplexed device,
it will be firstly diffracted by the R grating in all the possible orders, that will eventually enter the second grating.
Each of these beams are in turn recursively diffracted by the B grating, but only few of them possess the correct direction for further propagation (total internal reflection TIR can occur) to the detector.

Otherwise, when a blue monochromatic beam $\lambda_{B}$ is considered (case ii, for equal incidence angle) each diffracted order will possess a smaller diffraction angle $\beta$ (with respect of the previous case) due to the shorter wavelength, therefore it is possible that some overlapping between blue and red orders would occur since more diffraction orders have a direction that can potentially enter the detector.

\subsection{Issues in details: the efficiency depletion due to subsequent gratings}\label{sec:eff}
In the design of VPHGs for an astronomical spectrograph, after having satisfied the dispersion and resolution requirements, which fix parameters like the line density ($\Lambda$) of the gratings and the incidence and diffraction angles ($\alpha$ and $\beta$), the optimization of the diffraction efficiency (both peak efficiency and bandwidth) is necessary.

To perform this task, the main parameters to be considered are: 

\begin{enumerate}
\item the refractive index modulation $\Delta$n; 
\item the active film thickness \textit{d}; 
\item the slanting angle $\phi$ (i.e. the angle between the normal of the grating surface and the normal of the refractive index modulation plane).
\end{enumerate}

Considering a sinusoidal refractive index modulation, and working in the Bragg regime (the light is sent only in one diffraction order other than the zero), the well-known Kogelnik model can be used to compute the grating’s efficiency \citep{Kogelnik1969}. For small angles, large diffraction efficiency is achieved when the product $\Delta$n $\cdot$ $d$ is equal to half of the wavelength and this is the starting point in the optimization process. As already stressed, during the VPHG design, not only the peak efficiency is important, but also the efficiency at the edges of the spectral range. According to the Kogelnik model, the spectral bandwidth ($\Delta\lambda$) of the diffraction efficiency curve is proportional to \citep{Barden2000}:

\begin{equation}
\frac{\Delta\lambda}{\lambda} \propto \frac{\cot \alpha}{\Lambda d} 
\label{eq:koge_band}
\end{equation}

In this equation, $\alpha$ is the incidence angle, $\Lambda$ is the line density of the grating and it is evident that the bandwidth is inversely proportional to $\Lambda$ and the thickness of the grating $d$. Hence, the optimization of the diffraction efficiency curves, acting on the $\Delta$n and $d$, provides large differences in the grating response. 

If a grating works in the Bragg regime, the largest peak efficiency and bandwidth is obtained for very thin films and large $\Delta$n. Undoubtedly, the $\Delta$n upper value is determined by the performances of the holographic material. If the VPHG works in the Raman-Nath regime \citep{Moharam1978}, it diffracts the light with a non-negligible efficiency in more than one diffraction order and this is the case of low dispersion gratings and should be considered to avoid the further explained second order contamination 
(see Section \ref{sec:secondorder}).

For such gratings, the light diffracted in high orders is proportional to $\Delta$n$^2$, \textit{ergo} it is better to increase the film thickness and reduce the $\Delta$n in order to achieve a large peak efficiency. 
% Moreover, a VPHG working at longer wavelengths requires larger values of $\Delta$n and/or $d$ to achieve the same efficiency performances, both if it works in the Raman-Nath and Bragg regimes.

The availability of an holographic material that can exploit a precise ability to tune the $\Delta$n \citep{Zanutta2014, Zanutta2016}, is therefore crucial for the design of multiplexed elements, in order to be able to adjust the efficiency response of each dispersive layer.

\vspace{0.5cm}

%The most important issue to consider when designing a multiplexing device is how each single efficiency (coming from each diffracting layer) affects the subsequent one, taking into account every contribution that lowers the light transmitted through the entire device.

In the multipexing context, it is important to evaluate how a grating with a certain efficiency affects the response of the following one.
In order to give a feeling of the complexity of the problem, let us reconsider the two-multiplexing device in Figure \ref{fig:stack_scheme}. 
The total multiplexed efficiency on the detector will not merely be the sum of the single layer efficiencies $\eta_{B,1st}(\lambda, \alpha_0)$, $\eta_{R,1st}(\lambda, \alpha_0)$, with $\lambda$ the wavelength and $\alpha_0$ the initial incidence angle.

The notation "$_{1st}$" indicates the diffraction order at which the efficiency $\eta$ refers to, meaning that the system is aligned to work out the 1$^{st}$ order.

The spectrum generated by the R grating, before reaching the detector, has to pass through the B grating, and this will eventually diminish its intensity.
To complicate that, we add the fact that each wavelength of the R spectrum possess different diffraction angles $\beta_R$ (which became the incidence angles for the B grating) and therefore this varies the response from the second grating, according to the grating equation (eq. \ref{eq:grating}).
The resulting R efficiency $\eta^*_{R,1st}(\lambda, \alpha_0)$ will be then:

\begin{equation}
\eta^*_{R,1st}(\lambda, \alpha_0) = \eta_{B,0th}(\lambda, \beta_R) \cdot \eta_{R,1st}(\lambda, \alpha_0)
\end{equation}

%Note that $\beta_R$, wavelength dependent, is the diffraction angle of the R grating and the incidence angle for the B grating at the same time.

Moreover, the light that enters the second layer has already been processed by the previous gratings, therefore, its final efficiency will be the product of the leftover intensity, times the efficiency of the B grating:

\begin{equation}
\eta^*_{B,1st}(\lambda, \alpha_0) = \eta_{R,0th}(\lambda, \alpha_0) \cdot \eta_{B,1st}(\lambda, \alpha_0)
\end{equation}

%After having taken into account the contributions of all propagated orders, one can sum the efficiencies to retrieve the final throughput of the multiplexed element.
Practically, the goal is to obtain gratings with negligible overlapping efficiencies. 

\subsection{Issues in details: spectral range and Second Order Contamination}\label{sec:secondorder}
A critical point in spectroscopy is the contamination of recorded spectra, usually obtained through the first diffraction order, by light coming from other diffraction orders, usually the second.
Since signals of the different orders are overlapped, there is no possibility to remove the unwanted light \textit{a posteriori} with data reduction.
Therefore, dispersive elements with spectral range greater than $[ \lambda$ to $2 \lambda]$ will inevitably suffer of this problem.

This issue is well known in astronomy and it is usually avoided by placing order-sorting filters coupled with the dispersing element or in a filter wheel in the optical path of the instrument. The filters serve to block the light at lower wavelength that can overlap to the acquired spectrum. Another approach is to reduce as much as possible the efficiency of the unwanted orders.

Although in VPHGs, it is possible to mitigate these effect by varying parameters, such as thickness and $\Delta$n, in order to finely tune the efficiency curve, we decided to limit the wavelength range of the multiplexed device, adopting a spectral band where no contamination occurs.
However, we will show another strategy that deals with the second orders in the forthcoming "Part II" of this work.

\begin{figure}
\begin{center}
	\includegraphics[width=0.9 \columnwidth]{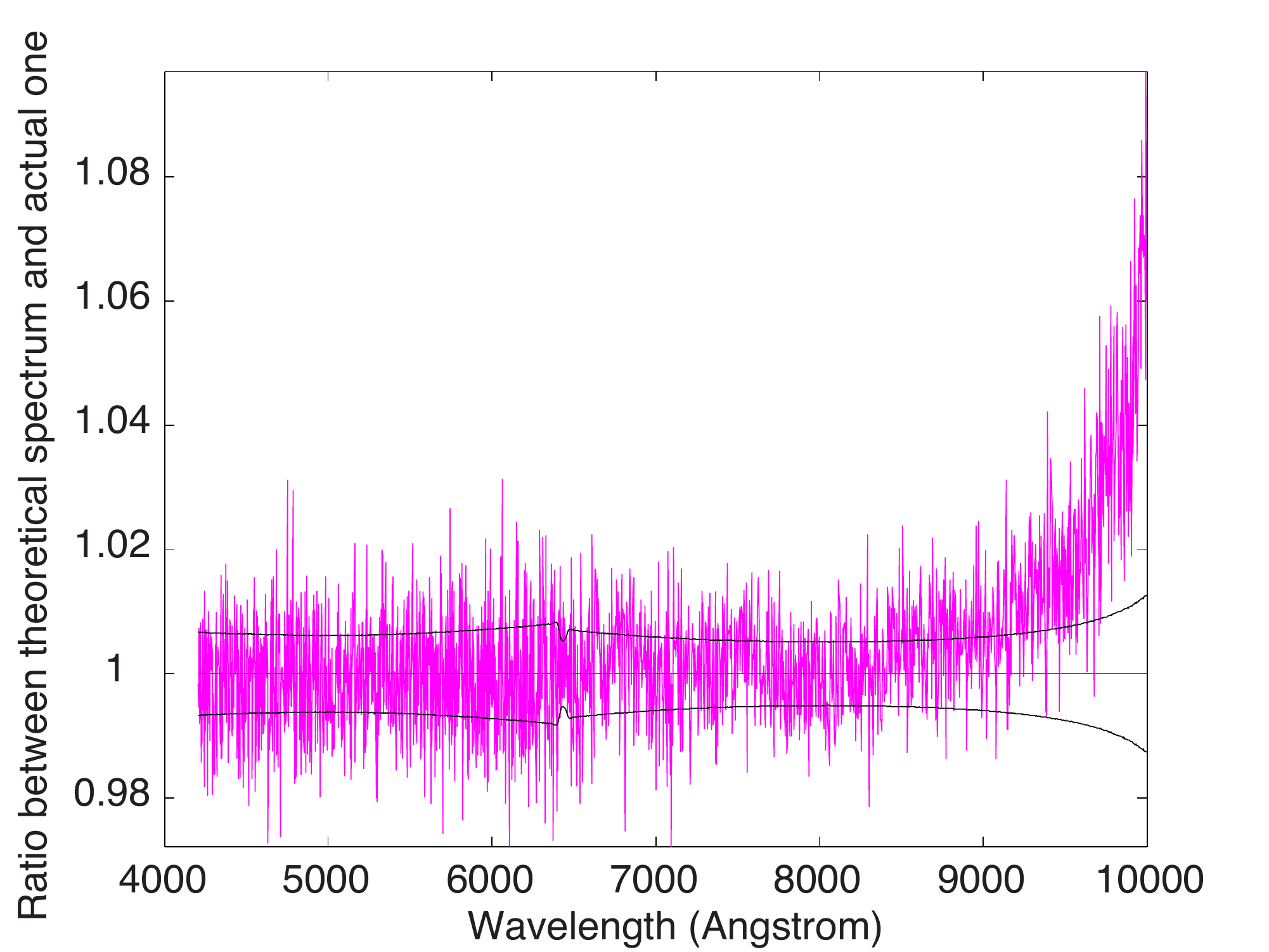}
    \caption{Ratio between contaminated and non-contaminated spectrum (magenta line) for a source with a power law spectral energy distribution at SNR of $\sim 100$. Black lines correspond to the detection limits dictated by SNR. The 2nd order contamination appears as a recognizable peak surpassing the black lines which corresponds to the noise level.}
    \label{fig:2_OSIRIS_SNR}
\end{center}
\end{figure}

In Figure \ref{fig:2_OSIRIS_SNR} we report the effects of second order contamination with a multiplexed device designed to work in an extended wavelength range [4200 - 10000 $\AA$]. We have simulated two different spectra: 
the first one contaminated with photons coming from the second order, while the other one considering only the contributions from the first order. We also assumed a SNR of about 100.
The ratio between the two signals (magenta line) is a rough estimation of how much  unwanted light appears as a bump in the collected data. In fact, according to this figure, the ratio between the two is within the noise level (solid black lines) up to $\lambda \simeq 9000 \AA$, indicating the two spectra are undistinguishable.
Surpassing this wavelength the ratio is higher than the noise level, meaning that photons from the second order are superimposing in the spectrum.

% \subsection{Issues in details: the effect of the tilt onto the efficiency}
% The last known effect that could damp the final efficiency of a stacked device is related to the tilt between the multiple gratings, necessary to separate the spectra into the detector. Its magnitude should be considered carefully, taking into account the minimum separation needed in the observations. As the grating rotates inside the GRISM device (grating within prisms), the incidence angle decreases (the projection should be considered), resulting in a change of performances. For the purposes of this work we considered a tilt of 2.5° between stacked gratings, since the effect on the efficiency is negligible.

\section{Design concepts and expected performances}\label{sec:simu}

\subsection{Theoretical framework}
In order to understand the spectral behavior of a multiplexing dispersive element, we choose to study the feasibility of this system considering an astronomical instrument that could take advantage of the multiplexing technology. 

%%%%%

The \textit{resolving power} of a spectrograph, R (or simply resolution) is:
\begin{equation}
R = \frac{m \Lambda \lambda W}{\chi D_{T}} 
%R = \frac{\lambda}{\delta \lambda} = \frac{m \Lambda \lambda D_{c}}{\cos (\beta) \chi D_{T}} 
\label{eq:resolving}
\end{equation}
where $W$ is the length of the illuminated area on the grating by the collimated beam, $\chi$ is the angular slit width (projected on the sky) and $D_{T}$ is the diameter of the telescope.
For a correct interpretation of the results, it has to be pondered that, the optical layout of the spectrograph (such as the ratio of telescope diameter and the collimated beam in the spectrograph) can be used as a rule of thumb to
quantify the advantage of this approach.
%%%%%

In this paper we present the case of the focal reducer OSIRIS, installed at the 10 m telescope Gran Telescopio Canarias, as candidate facility for the on-sky commissioning of the multiplexed device.
We have chosen to exploit two different case studies, changing the number of elements in the multiplexed device. A two-stacked multiplexed device with an approximate resolution of R $\sim$ 2000, and a three-stacked multiplexed device with a resolution of about R $\sim$ 5000.
The first one was intended to compete with observations carried out at the same facility for the determination of redshift lower limit of the TeV $\gamma$-ray BL Lacertae (BL Lac) object S4 0954+258 (see \cite{landonigtc} and the next sections) while the second one is intended to be compared with medium-high state-of-the-art resolution spectrographs such as ESO-XSHOOTER \citep{vernet11, lopez16}. In this section, we demonstrate the design and applicability of the two cases. For each of them, we present the analysis to determine the efficiency behavior, taking into account the dispersion and spectral range that each VPHG should show in relation to the instrument specifications. This activity is carried out both through optical ray-tracing and Rigorous Coupled Wave Analysis RCWA simulations \citep{moharam1981rigorous}. The outputs of this calculation are the most suitable efficiency curves for each stack that will guarantee the higher overall diffraction efficiency and are computed varying the key parameters described in Section \ref{sec:eff}. 

After the grating design, the subsequent step is to assess, through simulations, the expected on-sky performances of each device. Thus, we build up syntethic simulated spectra (starting from powerlaw model, as in the case of BL Lac, or template spectrum as in the case of QSO)  of the targets according to the expected signal-to-noise ratio (SNR) in each pixel defined as:

\begin{equation}
\frac{S}{N} = \frac{N_{*}}{\sqrt{N_{*} + N_{sky} + n_{pix} \cdot RON^2}} 
\label{eq:SNR}
\end{equation}
where $N_{*}$ is the number of expected counts from the target source evaluated as:

\begin{equation}
N_{*} = f(\lambda) \cdot \prod_{i=1}^n \cdot \eta_{i}(\lambda) \cdot A \cdot t_{exp}
\label{eq:Nstar}
\end{equation}
where $f(\lambda)$ is the input spectrum in ph sec$^{-1}$ cm$^{-2}$ $\textrm{\AA}^{-1}$, A is the collecting area of the telescope, $\eta_i$ are the efficiencies of atmosphere transmission, telescope, spectrograph, multiplexed device and CCD (we consider a slit efficiency of about $\sim$ 0.80 arcsec) and $t_{exp}$ is the total integration time. The quantity $N_{sky}$ is evaluated in the same way considering a flat spectrum normalized in V o R band to a flux that corresponds to $\sim$ 21 mag arcsec$^{-2}$, which is a typical value for the La Palma sky brightness. The read-out-noise (RON) of the detector is assumed equal to 7 e-/pix. For each simulation, we consider a seeing of $\sim$ 0.80 with a slit width of $\sim$ 0.60$^{\prime\prime}$ to be comparable with the current available instrumentation specifications and performances \citep{cepa10}. The plate scale of the system is assumed to be $\sim$ 0.30 arcsec/pix while the efficiencies of the telescope and optics of the spectrograph are derived from literature \citep{cepa10}.

%%%%%%%%%%%%%%%%%%%%%%%%%%%%%%%%%%%%%%%%%%%%%%%%%%%%

\subsection{Two-multiplexed grating case}
This first case that we took into consideration is a two stacked multiplexed device, for OSIRIS and in GRISM mode, that can cover in one single exposure a spectral range from 4000 to 8000 $\AA$ with a resolution of approximately 2000. 

\begin{figure}\begin{center}
	\includegraphics[width=0.55 \columnwidth]{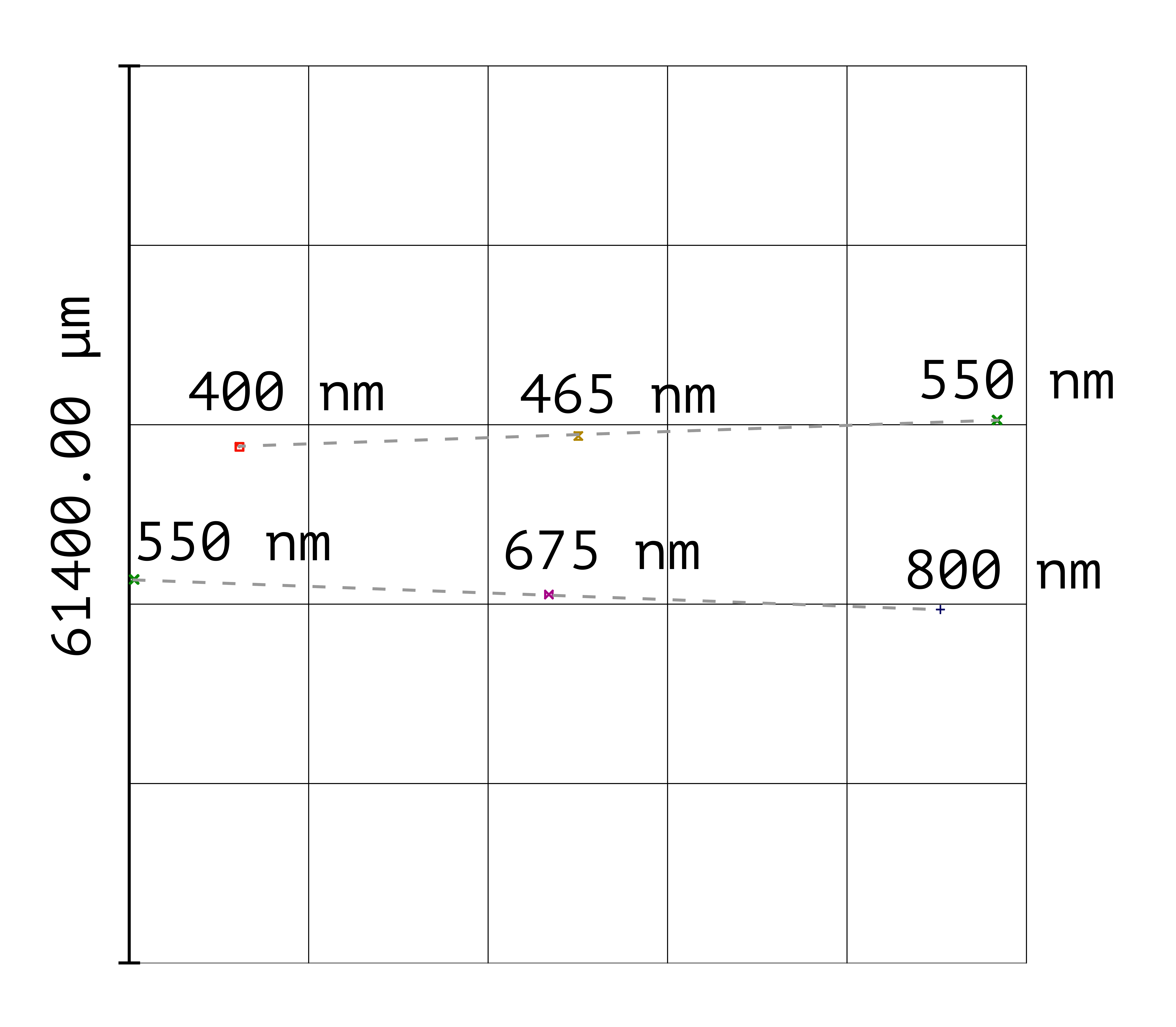}
    \caption{Simulated spectra onto the OSIRIS detector with the two layers multiplexed grating.}
    \label{fig:2_OSIRIS_detector}
\end{center}\end{figure}

In order to achieve that, the dispersive element splits the wavelength range in two parts, that are imaged on the detector one on top of the other. The inter-spectra-separation depends on the tilt of the two gratings in the diffraction element. In this particular scenario, the minimum distance between the two is approximately 2' (projected angle on the sky), but merely because we have chose arbitrarily a tilt value of 2.5° (see Figure \ref{fig:2_OSIRIS_detector}). The two dispersive elements share the same prisms and thus the same incidence angle.

\begin{figure}\begin{center}
	\includegraphics[width=0.9 \columnwidth]{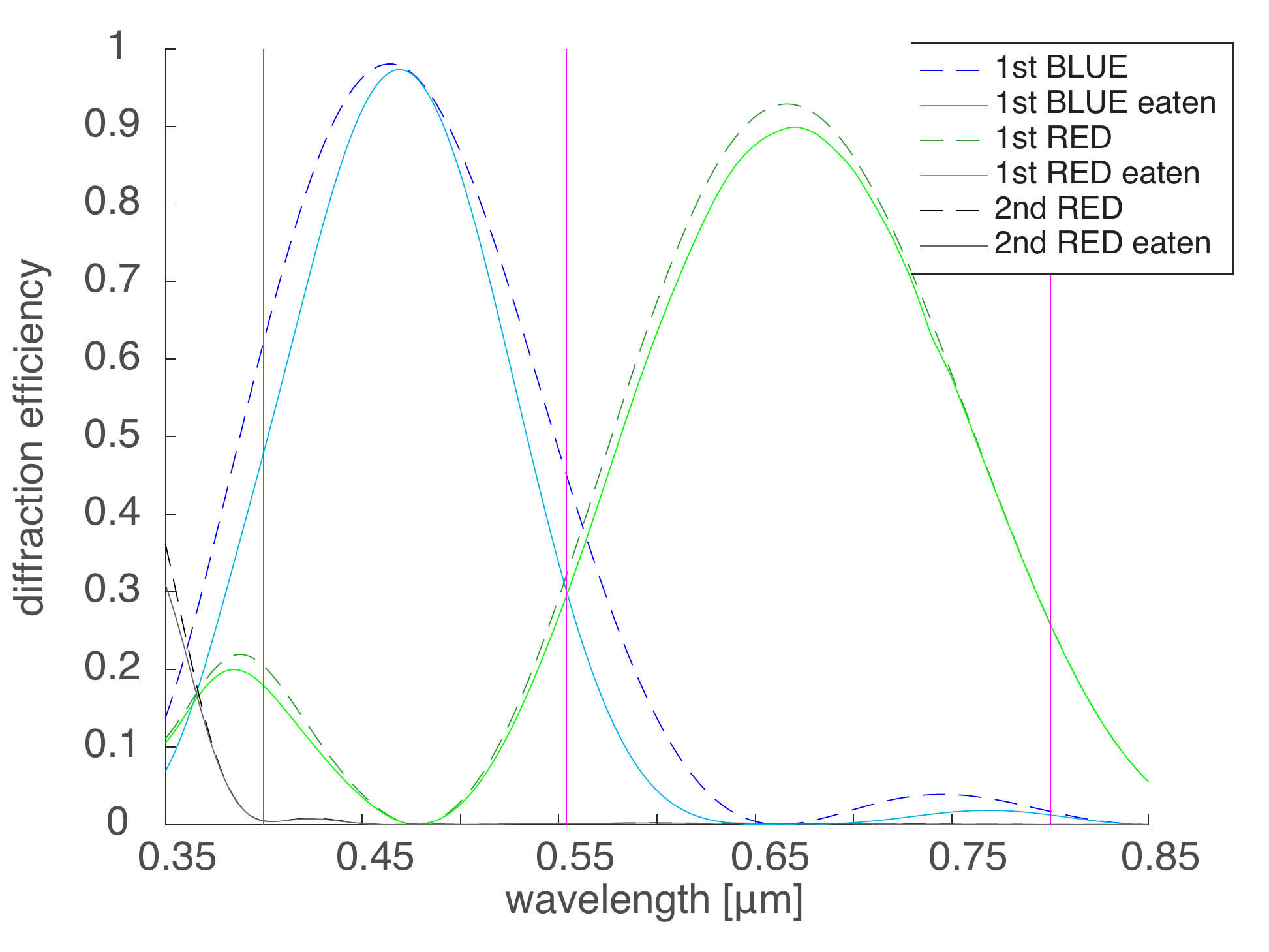}
    \caption{Diffraction efficiencies of the gratings composing the multiplexed element. The dotted lines refer to the single layer efficiencies (1st and 2nd diffraction orders), while the solid lines referto the corrected efficiencies (labelled "eaten") at the exit of the multiplexed element, due to the reciprocal interference of the dispersive layers. Vertical lines identify the wavelength boundaries of the spectra in the CCD for each VPHG. "Blue 2" is a VPHG with $\Delta$n = 0.038 and d = 6 $\mu$m while "Red 2" with $\Delta$n = 0.024, d = 12 $\mu$m and $\phi$ = 0.5$\mathrm{^o}$.}
    \label{fig:2_OSIRIS_complete}
\end{center}\end{figure}

In Table \ref{tab:osiris_2} the specifications of the gratings that have been designed are reported, while in Figure \ref{fig:2_OSIRIS_complete} we presented the calculated efficiency curves of the layers that compose the device. 
With respect of this last figure, a long-pass filter at 4000 $\AA$ is installed in the device in order to avoid contamination from the second order. Moreover the VPHG, which disperses the light in the range 5500 - 8000 $\AA$, has been designed to suppress as much as possible the contribution from the second order, which remains outside the spectral range.

As highlighted in the previous sections, an important effect that has to be taken into account is that the diffracted intensity will be dimmed as light gradually passes through the VPHG layers but, in this configuration, thanks to the precise design process, this effect is minimized.
Indeed, for each grating layer, a specific value of $\Delta$n, $d$ and slanting angle $\phi$ was chosen in order to ensure the compatibility between the efficiency curves.

In the hypothesis that the sequence is first the RED grating and second the BLUE grating, the wavelengths that are diffracted by the RED grating (dotted green in Figure \ref{fig:2_OSIRIS_complete}), are then transmitted through the BLUE layer with the resulting efficiency plotted in solid green. On the other hand, the wavelengths that have to be diffracted by the BLUE grating, must firstly pass though the RED layer, with a resulting efficiency that is plotted in solid blue.

\begin{figure}\begin{center}
	\includegraphics[width=0.9 \columnwidth]{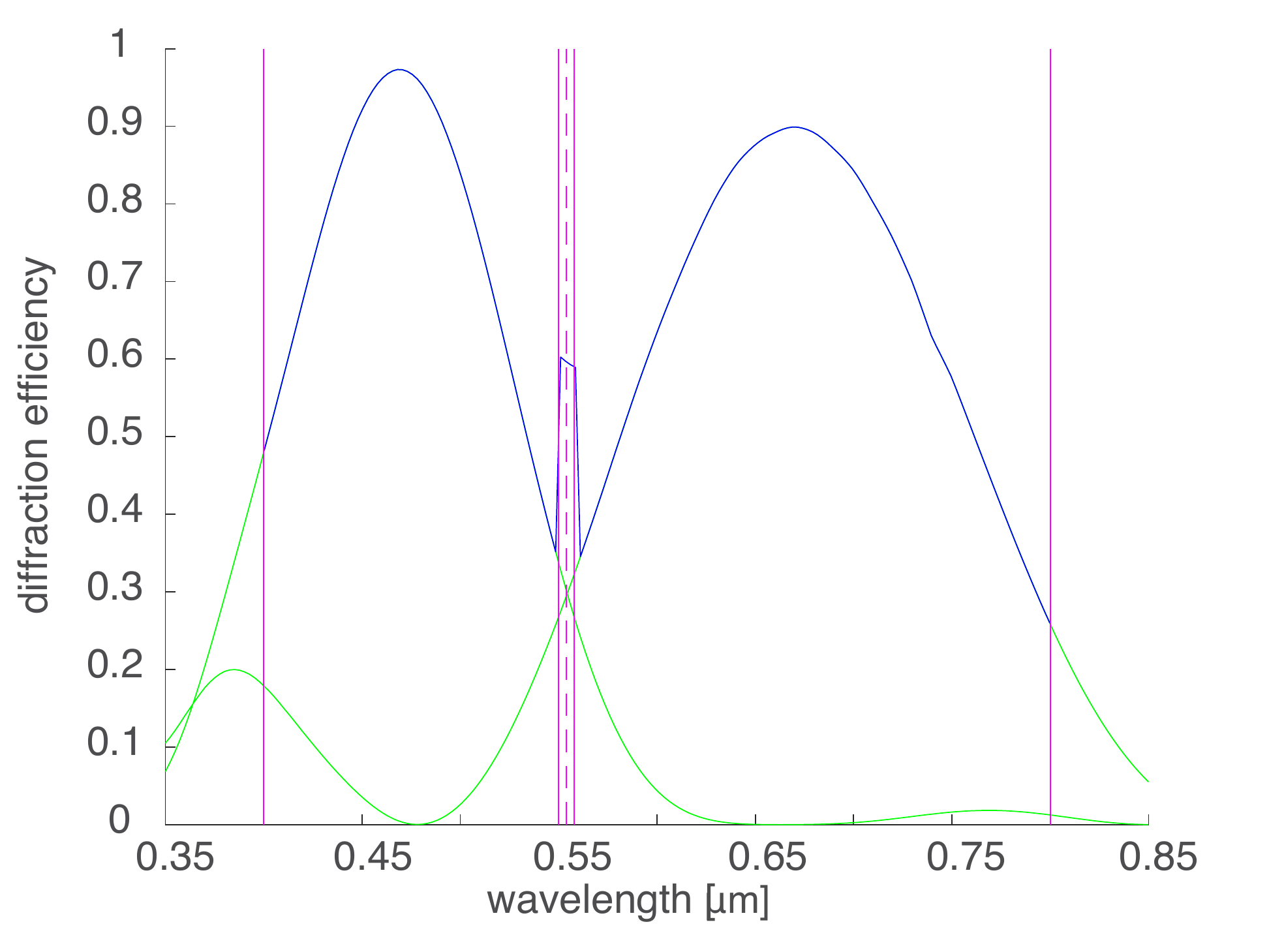}
    \caption{Blue line: overall efficiency of the two-multiplexing dispersive element. Green lines: single grating efficiencies of the spectra that are reaching the detector's focal plane. Vertical lines identify detector's boundaries.}
    \label{fig:2_OSIRIS_tot}
\end{center}\end{figure}

After accounting for all of these effects, the obtained efficiency curve for the multiplexed dispersive element is reported in Figure \ref{fig:2_OSIRIS_tot}.
The bump in the central region is due to light diffracted by both gratings and that falls on the detector in different places.
%shows the wavelength range that is duplicated in the detector and that can be used for orders cross-correlation during the data reduction.

Finally we remind that the efficiencies presented in the simulations do not take into consideration the material absorptions or the reflection losses that could arise to the presence of interfaces inside the device. Nevertheless we expect that these effects could be negligible at this level and are of the order of few percent points.

\begin{table}
	\centering
	\caption{Parameters of the stacked grating composing the two-multiplexed device for OSIRIS, with prisms' apex angle of 36.0°.}
	\label{tab:osiris_2}
	\begin{tabular}{lccccr} % four columns, alignment for each
		\hline
		grating & l/mm & $\lambda$ range  &  $\lambda_{centr.}$  &  R$_{0.6}$ & dispersion\\
			&	&	[nm]&	[nm]&	$@ \lambda_{central}$  &	 [\AA / px] \\
		\hline
		blue 2	& 	1500 & 	400-558 & 	475 &	2232 & 	0.52\\
		red 2	& 	1000 & 	550-800 & 	675 &	2086 & 	0.78\\
		\hline
	\end{tabular}
\end{table}

%% tabella delta enne, d e slanting angle e range
%\begin{table}
%	\centering
%	\caption{OSIRIS 2 grating parameters from MATLAB.}
%	\label{tab:osiris_2matlab}
%	\begin{tabular}{lccccr} % four columns, alignment for each
%		\hline
%		grating & l/mm & $\lambda$ range  & $\Delta$n &  d & slanting\\
%			&	&	[nm]&	&	[$\mu$m] &	 [$\mathrm{^o}$] \\
%		\hline
%		blue 2	& 	1500 & 	400-558 & 	0.038 &	6 & 	90\\
%		red 2	& 	1000 & 	550-800 & 	0.024 &	12 & 	90.5\\
%		\hline
%	\end{tabular}
%\end{table}

%%%%%%%%%%%%%%%%%%%%%%%%%%%%%%%%%%%%%%%%%%%%%%%%%%%%

\subsubsection{The case of S4 0954+65}

S4 0954$+$65 is a bright BL Lac object identified for the first time by \cite{walsh1984spectroscopy} which exhibits all the properties of its class. In particular, the source presents a strong variability in optical, with R apparent magnitudes usually ranging between 15 and 17 \citep{raiteri99}, linear polarisation \citep{morozova14} and a radio map that shows a complex jet-like structure. This BL Lac has recently caught attention since it was detected with the Cherenkov telescope MAGIC with a 5-$\sigma$ significance \citep{mirzoyan15}. The determination of redshift of BL Lac objects (in particular for TeV sources) is mandatory to assess their cosmological role and evolution, which appears to be controversial due to redshift incompleteness \citep{ajello14} and to properly understand their radiation mechanism and energetics (see e.g \cite{falomorev} and references therein). When BL Lacs are detected at TeV regime, the knowledge of their distance is unavoidable since they could be exploited as a probe of the Extragalactic Background Light (EBL, see e.g \cite{dominguez11}, \cite{franceschini08}) allowing to understand how extremely high energy photons propagate from the source to the Earth  and interact with the EBL through $\gamma$-$\gamma$ absorption. 
Unfortunately, the determination of the redshift of BL Lacs has proven to be difficult (see e.g. \cite{shaw13}, \cite{landonifors}, \cite{massarorev}) since their very faint spectral features are strongly diluted by their non-thermal emission (see the review of \cite{falomorev}). 

In the era of 10 m class telescopes (like the GTC), the research in this field has reached the so called "photon starvation regime" since the only way to significantly increase the SNR is the adoption of extremely large aperture telescope (like ELT) \citep{landonifors}.

On the other end, one can greatly increase the resolution of the secured spectra, maintaining a high SNR, decreasing the minimum Equivalent Width (EW$_{min}$), allowing to measure  fainter spectral features (see e.g. \cite{sbarufatti06}, \cite{shaw13}).

In particular, S4 0954$+$65 has been observed by \cite{landoni2015redshift} after its outburst on the night of February 28th, 2015. The object was observed with two grisms (R1000B and
R1000R) in order to ensure a spectral coverage from 4200 to $\sim$ 10000 $\textrm{\AA}$ adopting a slit of 1.00$^{\prime\prime}$ with a resolution of R $\sim$ 800. For each grism, the total integration time was 450s that corresponds to about 0.5 hrs of telescope allocation time (including overheads). The collected data allowed to disprove previous redshift claims of $z = 0.367$ \citep{stickel93, lawrence96} and to infer a lower limit to the distance of $z \geq 0.45$ thanks to EW$_{min} \sim 0.15 \textrm{\AA}$ and SNR $>$ 100.

%\begin{figure}
%\begin{center}
%	% To include a figure from a file named example.*
%	% Allowable file formats are eps or ps if compiling using latex
%	% or pdf, png, jpg if compiling using pdflatex
%	\includegraphics[width=0.85 \columnwidth]{TH_osiris2sym.eps}
%    \caption{Throughput comparison between the multiplexed case and the standard OSIRIS grism R1000B.}
%    \label{fig:TH_osiris2sym}
%\end{center}
%\end{figure}

\begin{figure*}\begin{center}
	% croppata a 260mm di larghezza con punto centrale in AI
	\includegraphics[width=1.4 \columnwidth]{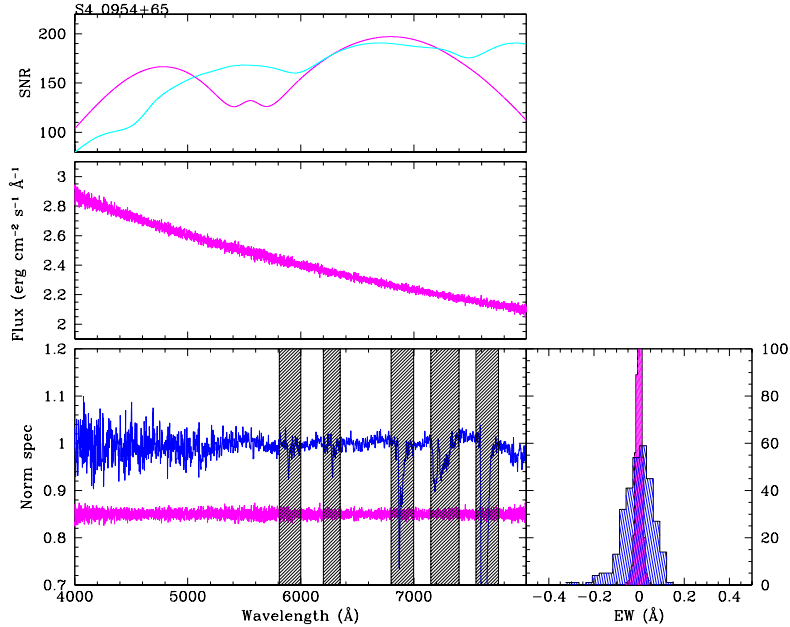}
        \caption{Two-Multiplexed grating case: Simulation of the S4 0954+65 spectrum (magenta) and comparison with the real observed spectrum (solid blue) secured with R1000B+R GRISMS. Spectral regions where telluric absorptions are severe are shaded and not included in the analysis. Bottom right box reports the comparison of the histograms of EW$_{min}$ between the observed spectrum in feb. 2015 (shaded blue) and the one obtained with the new dispersing device. Top left box reports the SNR of the spectrum of S4 0954+65 obtained with the new device (magenta) and the one estimated with GRISMs R2500 (cyan). }
    \label{fig:Land_1}
\end{center}\end{figure*}

In order to further increase the lower limit to the redshift or, even better, detect faint spectral features arising from the host galaxies that harbors this BL-Lac, the only straightforward solution with this state-of-the-art instrumentation is to drastically increase the resolution of the collected spectrum.

Considering the case of GTC and OSIRIS, the only available opportunity is to observe the target with the GRISMs R2500. Unfortunately these gratings possess a very narrow spectral range so in order to ensure the wavelength coverage similar to the required (4000 - 8000 $\textrm{\AA}$), one must collect four different spectra. This turns out in a telescope allocation time of about 2 hrs (including overheads).

By the adoption of the two VPHG multiplexed device, the observer is able to collect simultaneously two spectra with a whole spectral range from 4000 to 8000 $\textrm{\AA}$ with a resolution of approximately 2000.
The simulated spectra obtained with this device is reported in Figure \ref{fig:Land_1} along with the comparison of the R1000B+R observed one.

We also report the distribution of minimum detectable Equivalent Width, estimated following the recipes detailed in \cite{sbarufatti05} (histogram in the bottom right corner of Figure \ref{fig:Land_1}).

The detectable EW$_{min}$ on the spectrum simulated by assuming the new dispersing element is 0.03, which is a factor of 5 lower than the compared one.
This turns in a lower limit to the redshift of $z \gtrsim 0.55$ putting the source at a plausible redshift region where the absorption from the EBL becomes severe and making this TeV object an excellent probe for the study of the EBL through absorption.

In this figure we also report the expected SNR obtained with our device (solid magenta in the top left box), and the one simulated assuming the currently available R2500 devices at the GTC.

\subsection{Three-multiplexed grating case}
For the science cases that require a wide spectral range with a moderate resolution, nowadays the only possibility to fulfill the requirements is to adopt an echelle grating based instrument, which is capable to secure wide wavelength ranges in a reasonable number of shots .
Otherwise according to OSIRIS GRISMs specifications, in the GTC manual, up to six different setups (and exposures) are required to obtain the same result just in terms of spectral range, since the maximum resolution is approximately $R_{max}=2500$.

In this section we present a possible application of multiplexed VPHGs, aiming to refurbish the dispersive elements of OSIRIS at GTC, in order to reach the closest possible performance with respect to UV and VIS arm of X-SHOOTER. 

In order to cover a wide spectral range with a resolution of approximately 4500, we have designed two multiplexed dispersive elements, each one composed by three stacked layers, therefore they will produce on the detector three spectra for each single exposure. 
With these two devices together, in just two exposures, we can cover a range from 3500 to 10000 $\AA$. 

%This solution could allow to refurbish this instrument in such a way that it could be comparable with a medium resolution, high throughput spectrograph, e.g. X-SHOOTER \citep{vernet11}. 

While the number of dispersive layers could be theoretically further increased, due to complexities in calculations, possible transparency issues and manufacturing alignment, in this work we decided to set the limit to three elements.

\subsubsection{BLUE device, from 350 to 600 nm}
\begin{figure}\begin{center}
	\includegraphics[width=0.9 \columnwidth]{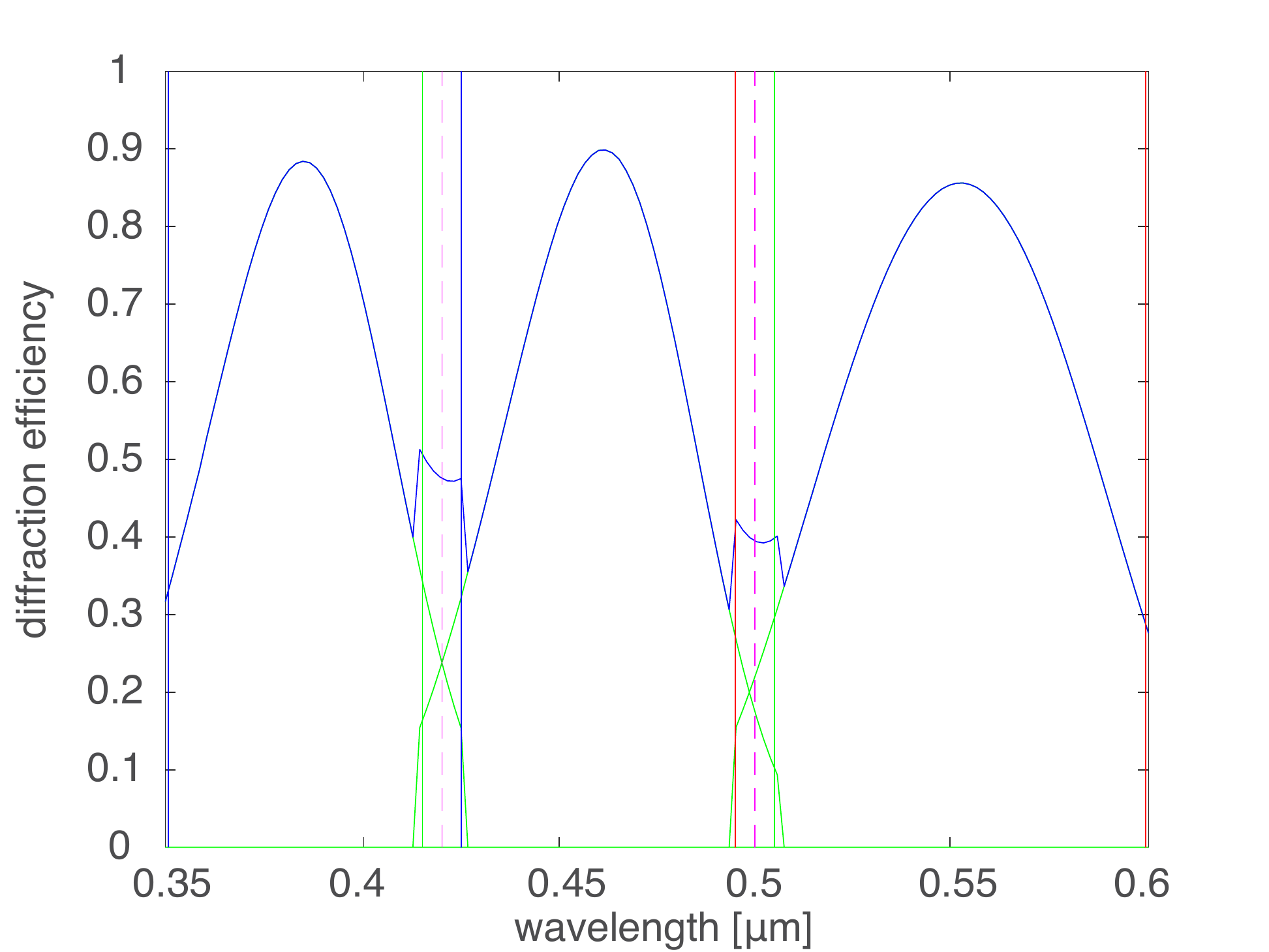}
    \caption{Multiplexing dispersive element designed for the BLUE band. Blue line: overall efficiency, green lines: single layer efficiencies of the VPHGs. In the same way explained for the two-multipexed case, vertical lines identify detector's boundaries. "Blue 3.1" possess $\Delta$n = 0.055, d = 4 $\mu$m, "Blue 3.2" $\Delta$n = 0.037, d = 6 $\mu$m and $\phi$ = -0.2$\mathrm{^o}$, "Blue 3.3" $\Delta$n = 0.035, d = 7.5 $\mu$m.}
    \label{fig:3_OSIRIS_totB}
\end{center}\end{figure}

The first (of two) multiplexed device will be responsible for the dispersion of the light in the range  3500 -- 6000 $\AA$, therefore hereafter it will be identified as the BLUE device.
It is composed by three dispersing layers, each of them generating the peaks in the summed efficiency displayed in Figure \ref{fig:3_OSIRIS_totB} (solid blue curve). For this case, we did not report the plot with the contributions that generate the overall efficiency, since the general procedure is the same that in the two-multiplexed case.
As highlighted in the previous case, the vertical solid lines identify the size of the detector with respect to each spectra: since the total range will appear divided in three parts, the upper is displayed with solid blue boundaries, the central with green and the lower with red.
As some small portions of the range will overlap, bumps in efficiency in the regions between the peaks  appear.

In Table \ref{tab:osiris_3B}, we report the specifications of the three gratings that have been designed for this BLUE element along with the calculated resolution and dispersion that is achievable integrating this device in the OSIRIS spectrograph.

\begin{table}
	\centering
	\caption{Parameters of the BLUE stacked grating composing the three multiplexed device for OSIRIS, with prisms' apex angle of 52.3°.}
	\label{tab:osiris_3B}
	\begin{tabular}{lccccr} % four columns, alignment for each
		\hline
		grating & l/mm & $\lambda_{centr.}$  &  $\lambda$ range &  R$_{0.6}$ & dispersion\\
			&	&	[nm]&	[nm]&	$@ \lambda_{central}$  &	 [\AA / px] \\
		\hline
		BLUE 3.1	& 	2850 & 	385 & 	354-425 &	4339 & 	0.24\\
		BLUE 3.2	& 	2400 & 	460 & 	411-498 &	4430 & 	0.29\\
		BLUE 3.3	& 	1980 & 	550 & 	493-600 &	4346 & 	0.35\\
		\hline
	\end{tabular}
\end{table}

%\begin{figure}\begin{center}
%	% To include a figure from a file named example.*
%	% Allowable file formats are eps or ps if compiling using latex
%	% or pdf, png, jpg if compiling using pdflatex
%	\includegraphics[width=0.85 \columnwidth]{multiplexing_3_OSIRIS_UV_inv_complete}
%    \caption{Diffraction efficiencies of the gratings composing the multiplexed element in function of the wavelength. The "EATEN" labelled curves are calculated taking into account that the disposition of the various layers causes a decrease in the single grating efficiencies. Vertical lines identify the wavelength ranges of the different VPHGs.}
%    \label{fig:3_OSIRIS_complete}
%\end{center}\end{figure}

%% tabella delta enne, d e slanting angle e range
%\begin{table}
%	\centering
%	\caption{OSIRIS 3 grating parameters from MATLAB.}
%	\label{tab:osiris_3matlab}
%	\begin{tabular}{lccccr} % four columns, alignment for each
%		\hline
%		grating & l/mm & $\lambda$ range  & $\Delta$n &  d & slanting\\
%			&	&	[nm]&	&	[$\mu$m] &	 [$\mathrm{^o}$] \\
%		\hline
%		blue 3	& 	2850 & 	354-425 & 	0.055 &	4 & 	90\\
%		green 3	& 	2400 & 	411-498 & 	0.037 &	6 & 	89.8\\
%		red 3& 	1980 & 	493-600 & 	0.035 &	7.5 & 	90\\
%		\hline
%	\end{tabular}
%\end{table}

\subsubsection{RED device, from 600 to 1000 nm}
\begin{figure}\begin{center}
	\includegraphics[width=0.9 \columnwidth]{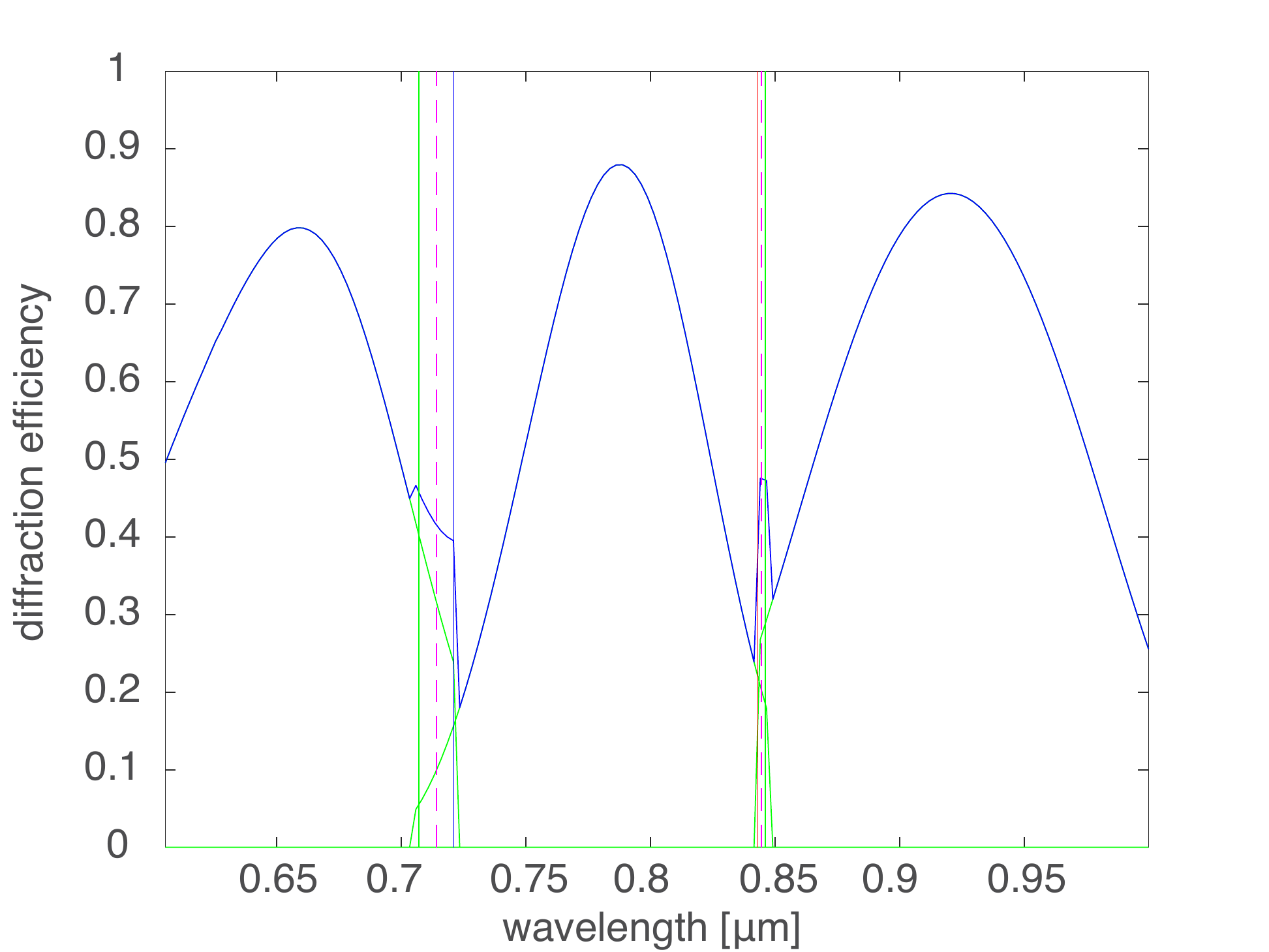}
    \caption{Multiplexing dispersive element designed for the RED band. Blue line: overall efficiency, green lines: single layer efficiencies of the VPHGs. In the same way explained for the two-multipexed case, vertical lines identify detector's boundaries. "Red 3.1" possess $\Delta$n = 0.055, d = 6 $\mu$m, "Red 3.2" $\Delta$n = 0.044, d = 10 $\mu$m and $\phi$ = -0.5$\mathrm{^o}$, "Red 3.3" $\Delta$n = 0.038, d = 12 $\mu$m.}
    \label{fig:3_OSIRIS_totR}
\end{center}\end{figure}

This second multiplexed device will be responsible to disperse the light in the spectral range from 6000 to 10000 $\AA$, therefore hereafter it will be identified as the RED device.
Figure \ref{fig:3_OSIRIS_totR} (solid blue curve) reports the overall efficiency curve that can be produced by the three dispersing layers composing this device.

In Table \ref{tab:osiris_3R}, we report the specifications of the gratings that have been designed for this RED element along with the calculated resolution and dispersion that is achievable integrating this device in the OSIRIS spectrograph.

\begin{table}
	\centering
	\caption{Parameters of the RED stacked grating composing the three multiplexed device for OSIRIS, with prisms' apex angle of 55.1°.}
	\label{tab:osiris_3R}
	\begin{tabular}{lccccr} % four columns, alignment for each
		\hline
		grating & l/mm & $\lambda_{centr.}$  &  $\lambda$ range &  R$_{0.6}$ & dispersion\\
			&	&	[nm]&	[$\mathrm{^o}$]&	$@ \lambda_{central}$  &	 [\AA / px] \\
		\hline
		RED 3.1	& 	1750 & 	655 & 	605-721 &	4825 & 	0.39\\
		RED 3.2	& 	1480 & 	775 & 	707-846 &	4851 & 	0.46\\
		RED 3.3	& 	1240 & 	920 & 	843-1000 &	4814 & 	0.55\\
		\hline
	\end{tabular}
\end{table}

%\begin{figure}\begin{center}
%	% To include a figure from a file named example.*
%	% Allowable file formats are eps or ps if compiling using latex
%	% or pdf, png, jpg if compiling using pdflatex
%	\includegraphics[width=0.85 \columnwidth]{multiplexing_3_OSIRIS_NIR_inv_complete}
%    \caption{Diffraction efficiencies of the gratings composing the multiplexed element in function of the wavelength. The "EATEN" labelled curves are calculated taking into account that the disposition of the various layers causes a decrease in the single grating efficiencies. Vertical lines identify the wavelength ranges of the different VPHGs.}
%    \label{fig:3_OSIRIS_complete}
%\end{center}\end{figure}

%% tabella delta enne, d e slanting angle e range
%\begin{table}
%	\centering
%	\caption{OSIRIS 3 grating parameters from MATLAB.}
%	\label{tab:osiris_3matlab}
%	\begin{tabular}{lccccr} % four columns, alignment for each
%		\hline
%		grating & l/mm & $\lambda$ range  & $\Delta$n &  d & slanting\\
%			&	&	[nm]&	&	[$\mu$m] &	 [$\mathrm{^o}$] \\
%		\hline
%		blue 3	& 	1750 & 	605-721 & 	0.055 &	6 & 	90\\
%		green 3	& 	1480 & 	707-846 & 	0.044 &	10 & 	89.5\\
%		red 3& 	1240 & 	843-1000 & 	0.038 &	12 & 	90\\
%		\hline
%	\end{tabular}
%\end{table}

%\clearpage

\subsubsection{Application to Extragalactic Astrophysics. The characterisation of Intergalactic medium}
The study of the Intergalactic and Circumgalactic medium (IGM and CGM) is a powerful tool to investigate the properties of the cool (and clumpy) gaseous halos between the observer and the source, that lies at a certain $z$. The only way to investigate the IGM or the CGM is through absorption lines imprinted in the spectra of distant QSO, as demonstrated in the last few years by e.g \cite{prochaska14, landoniciv, lopez16} since its surface brightness is extremely faint to be probed directly, and only few examples are know to succeed in the detection of emission of Ly-$\alpha$ lines in the CGM (e.g \cite{arrigoni15}). 
This research field is actively growing and, recently, has begun to probe not only the physical state and the chemical composition of the IGM but also the three-dimensional distribution of the gas allowing scientists to build up an actual tomography of the cool Universe between background quasars and the Earth. For example, in this context one of the most recent and successfully survey is the CLAMATO survey \citep{lee16}. In this projects, authors aims to to collect spectra for 500 background Lyman-Break galaxies (LBGs) in $\sim$ 1 sq degree area to reconstruct a 3D map with an equivalent volume of (100 h$^{-1}$Mpc)$^{3}$.

\begin{figure*}
\begin{center}
	\includegraphics[width=1.6 \columnwidth]{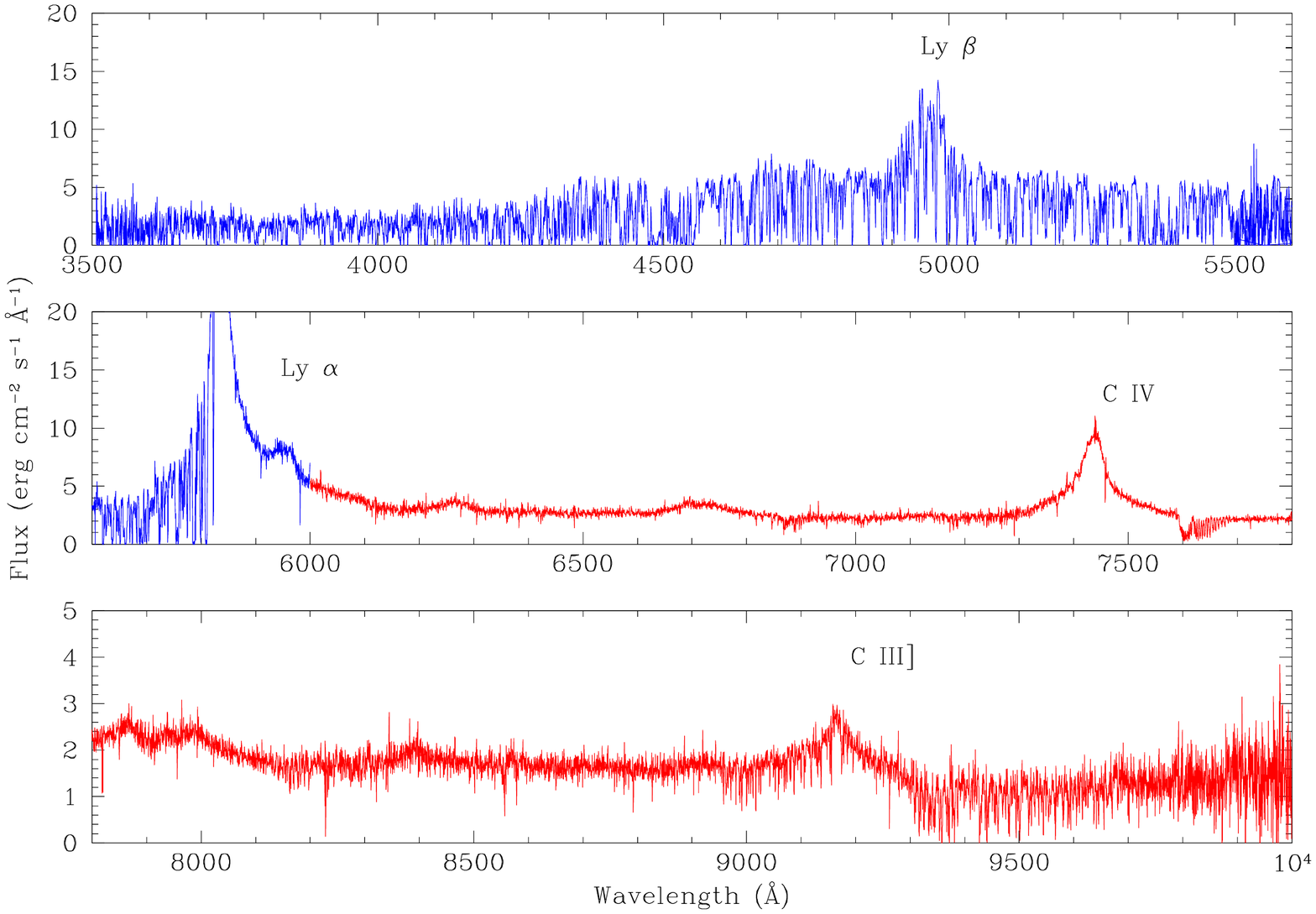}
    \caption{Simulation of a quasar spectrum at z $\sim$ 3.75, observed with our double Three-Multiplexed system. The solid blue line corresponds to the expected emission spectrum of the quasar recorded by the BLUE multiplexed device, it is worth noted that absorption lines from intervening systems are clearly visible and resolved in most cases. The solid red line corresponds to the part of the spectrum recorded with the RED multiplexed device }
    \label{fig:QSO}
\end{center}
\end{figure*}

The key step in these spectroscopic studies is the availability of moderate-high resolving power (R $\sim$ 5000) and wide spectral coverage, in order to probe as much as possibile absorption lines, perform diagnostic ratio to probe the interplay between galaxies and the intergalactic medium (IGM). 
However, such observations are typically time consuming and require good SNR at moderate R, a particularly high challenge for distant QSOs which tend to be faint.
For all of these reason, the availability of new instrumentation available to collect spectra in a wide spectral range (3500-1 $\mu$m) at R > 4000 would be really advantageous allowing to further increase the availability of telescopes able to tackle these kind of surveys, especially for facilities with moderate telescope aperture.
In order to demonstrate the applicability of our new device, we simulated the expected performance by assuming to observe for $t_{exp} = 200s$ for each grating a QSO (template taken from \cite{lopez16}) at redshift z = 3.78 with m$_{R} \sim 17$. The overall obtained spectrum is reported in Figure \ref{fig:QSO}. In particular the solid blue line corresponds to emission spectrum of  the quasar secured with the BLUE multiplexed device. The absorption lines, imprinted by Lyman-$\alpha$ intervening systems and used to probe the IGM, are clearly detected and resolved in most of the cases. The solid red line, instead, report the spectrum recorded with the RED multiplexed device where emission line from  C IV and C III] are visibile. Results reported in Figure \ref{fig:QSO} are obtained with a total integration time of about 400s while, by comparison, to obtain the same results at half resolution with grisms available at GTC-OSIRIS would require more than 1000s, since it should be observed four times with 4 different gratings.

As highlighted in the previous paragraphs, the X-SHOOTER spectrograph is able to obtain similar results with a broader band in a single snapshot. Although this outcome is obviously outside the capabilities of our proposed solution, the multiplexing VPHG allows to cover in just two snapshot a comparable quality (in terms of R, SNR and spectral range) in the UV and visible band.
Therefore, the integration of such element in a facility like OSIRIS would allow to scientifically compete with key-science projects that require spectroscopic capabilities otherwise available only with major instrument commissioning.

\section{Conclusions}\label{sec:conclusions}
We have demonstrated the theoretical feasibility and the advantages of an innovative dispersive element, able to greatly increase the performances of the existing spectrograph at the state of the art 10 m telescope GTC.
Thanks to the advantages derived by the adoption of the photopolymeric material considered in the simulations, we achieved to increase by at least a factor of two in terms of resolution (and thus in the spent observing time), without changes in the optical layout of the spectroscopic instrument.
We have also shown that in the case of the three-multiplexed VPHG, it is possible to reach with GTC OSIRIS, approximately the performances of the UV and VIS arm of X-SHOOTER (when operating in medium resolution) in just two exposures of the same target.
Even though in this work we have selected GTC OSIRIS for the simulations, the philosophy behind this multiplexing design could be applied to almost every focal reducer spectrograph, donating the discussed advantages to all the instruments, allowing them to handle scientific cases that would be otherwise out of reach for these facilities.
In the forthcoming second part of this work, we will realize and integrate the multiplexed device in a spectrograph for science verification, focusing on the observational cases highlighted in the simulations in this paper.

\section*{Acknowledgements}

This work was partly supported by the European Community (FP7) through the OPTICON project (Optical Infrared Co-ordination Network for astronomy) and by the INAF through the TECNO-INAF 2014 “Innovative tools for high resolution and infrared spectroscopy based on non-standard volume phase holographic gratings”.

\bibliographystyle{mnras}
\bibliography{references} % if your bibtex file is called references.bib

%%%%%%%%%%%%%%%%%%%%%%%%%%%%%%%%%%%%%%%%%%%%%%%%%%

%%%%%%%%%%%%%%%%% APPENDICES %%%%%%%%%%%%%%%%%%%%%

%\appendix



%%%%%%%%%%%%%%%%%%%%%%%%%%%%%%%%%%%%%%%%%%%%%%%%%%

% Don't change these lines
\bsp	% typesetting comment
\label{lastpage}
\end{document}